\documentclass[letter]{aa} 
\bibpunct{(}{)}{;}{a}{}{,} 
\usepackage{graphicx}
\usepackage{natbib}

\usepackage{txfonts}
\usepackage{hyperref}
\usepackage[normalem]{ulem}
\usepackage{subcaption}

\makeatletter
\let\linenumbers\relax
\makeatother

\begin{document}

\title{Hydrodynamical modeling of SN~2025kg associated with the Fast X-ray Transient EP250108a}

\titlerunning{Supernova 2025kg associated with an X-Ray Flash}

\author{L.M. Roman Aguilar\inst{1,2} \and M.C. Bersten\inst{1,2,3}
}
          
    \institute{Facultad de Ciencias Astronómicas y Geofísicas, Universidad Nacional de La Plata, Paseo del Bosque S/N B1900FWA, La Plata, Argentina\label{inst1} \and Instituto de Astrofísica de La Plata, CONICET, Argentina\label{inst2} \and 
    Kavli IPMU (WPI), UTIAS, The University of Tokyo, Kashiwa, Chiba 277-8583, Japan\label{inst3}}

   \date{Received /
   Accepted}

 \abstract
 {Supernovae (SNe) associated with X-Ray flashes (XRFs) are extremely rare. Therefore, the discovery of each new object in this class offers a unique opportunity to improve our understanding of their origins and potential connection with other high-energy phenomena. SN 2025kg is one of the most recent events discovered in this category and exhibits a double-peaked light curve, with an initial cooling phase followed by the main peak. Here, we investigate the possible mechanisms powering its bolometric light curve and expansion velocities, using numerical calculations to simulate the explosion. We find that low ejecta masses (M$_{\rm{ej}} \sim 2$ M{$_\odot$}) and moderate explosion energies (E $\sim 2\times 10^{51}$ erg) are required to reproduce the data. Our models also show that a  large amount of nickel (M$_{\rm{Ni}}\sim 0.85\, \rm{M}_\odot$) is needed to achieve the high luminosity of SN 2025kg, which makes this scenario difficult to sustain. As an alternative, we explore a model in which a millisecond magnetar serves as the primary energy source. A magnetar with a spin period of $\sim 3$ ms and a magnetic field of 28 $\, \times \, 10^{14}\,$G provides an adequate match to the data. 
 To account for the early cooling phase, we assume the presence of a dense circumstellar material surrounding the progenitor, with a mass of 0.27 M$_\odot$ and an extension of 500 R$_\odot$. A comparison and modeling of a selected group of SNe—SN~2006aj, SN~2020bvc, and SN~2023pel—is also presented. A remarkable similarity emerges between SN~2025kg and SN~2023pel. Since SN~2023pel was recently proposed to be powered by a magnetar, this further supports the magnetar scenario for SN~2025kg.
}


\maketitle
%

\section{Introduction}

\defcitealias{Gokul:25}{S25}
\defcitealias{Eyles-Ferris:25}{EF25}
\defcitealias{Rastinejad:25}{R25}
\defcitealias{Roman:2025}{RA25}
\defcitealias{Li:25}{L25}

The connection between long gamma-ray bursts (GRBs) and core-collapse supernovae (CCSNe) is now well established {(see \cite{Cano:2017} for a review)}. Numerous associations have been confirmed, with all corresponding to SNe classified as hydrogen- and helium-deficient objects, with broad spectral lines indicative of large kinetic energies (SNe Ic-BL). Although there are several confirmations, these explosions are still very rare compared to other types of CCSNe. Even less frequent are those SNe accompanying X-Ray flashes  (XRFs\footnote{Also known as Fast X-ray Transients, FXTs.}). The first clear identification of these events was SN~2006aj \citep{Soderberg:2006,Pian:2006}, followed a few years later by SN~2010bh \citep{Cano:2011,Olivares:2012}. Since then, no further associations have been reported until very recently. Given the scarcity of these events, every new object of this type deserves detailed study, as it may help to better understand the connections between XRF-SNe, GRB-SNe, and SNe Ic-BL. 

Recently, a new XRF-SN association has been reported \citep{Li:25}. The FXT (EP250108a) was discovered by the EP mission on 8 January 2025, and subsequent observations confirmed the existence of an optical counterpart, designated as SN~2025kg and classified as SN Ic-BL \citep{Eyles-Ferris:25}. Since this discovery, several works have been published presenting and analyzing the photometric and spectroscopic properties of this event \citet{Rastinejad:25} (R25), \citet{Li:25} (L25), \citet{Gokul:25} (S25), and \citet{Eyles-Ferris:25} (EF25).

The light curve (LC) of the SN 2025kg  was analyzed in \citetalias{Rastinejad:25}, \citetalias{Gokul:25} and \citetalias{Li:25}. The first two studies modeled the main emission using an analytical radioactive-decay model and derived nickel masses of 0.2–0.6 M$_\odot$ and $0.57^{+0.6}_{-0.3}$ M$_\odot$, respectively. While \citetalias{Li:25}
noted that SN~2025kg was brighter than other He-deficient SNe and proposed a magnetar as an additional power source to enhance its luminosity. The XRF properties and early data of SN~2025kg \mbox{(t < 6 days)} were presented in \citetalias{Eyles-Ferris:25,Gokul:25}, and \citetalias{Li:25}.
Several scenarios were explored for the initial optical emission; however, \citetalias{Gokul:25} and \citetalias{Eyles-Ferris:25} favored models incorporating an extended CSM to explain the early-time emission.
Previous works analyzed the properties of SN~2025kg using analytical prescriptions; here, we modeled the LC and expansion velocities using numerical simulations, based on public data. Our results are compared with other objects and previous studies, providing a broader context for SN 2025kg.

\begin{figure}
    \centering
    \includegraphics[width=0.95\linewidth]{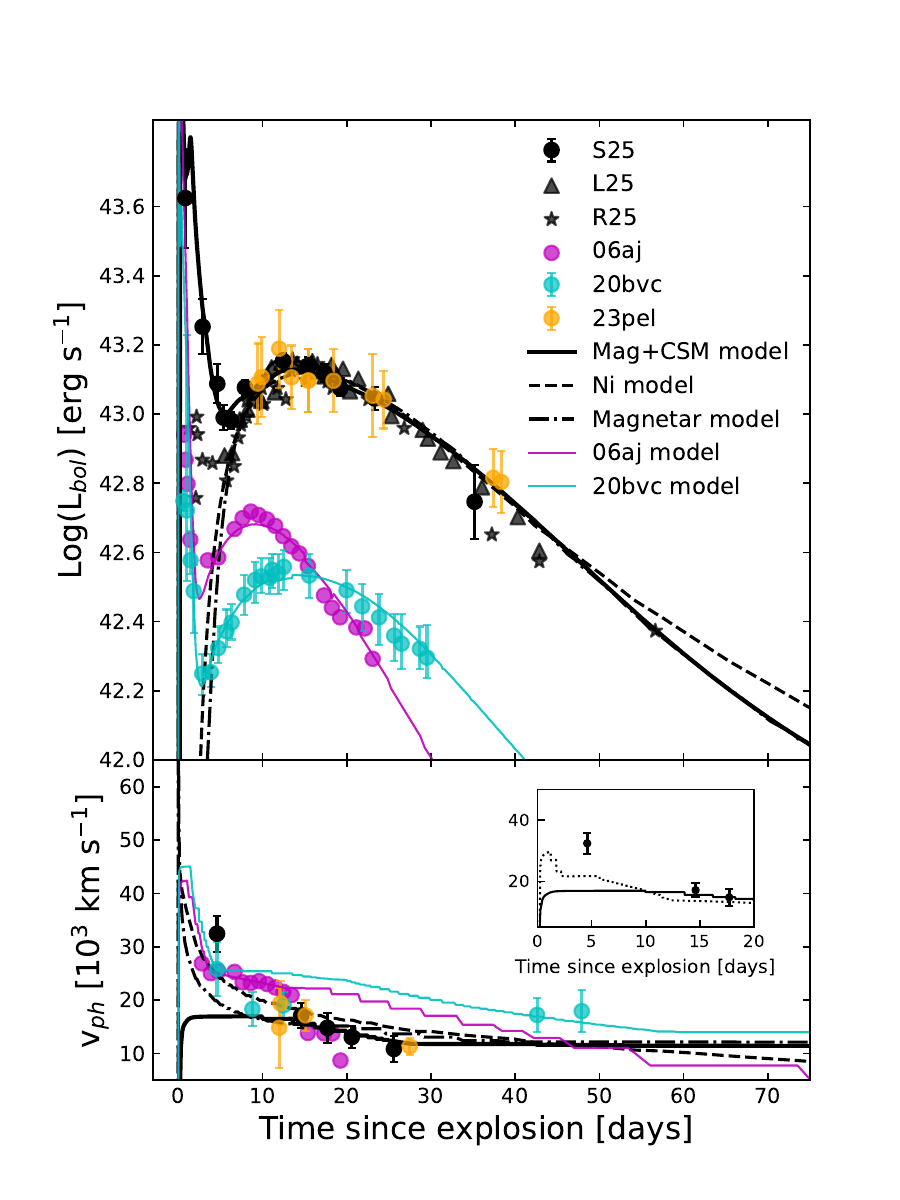}
    \caption{Comparison between SN~2025kg and a set of SNe associated with high-energy emission. Top panel: Bolometric LCs. Bottom panel: Photospheric and FeII line velocities. Black symbols show the available data for SN~2025kg (stars from \citetalias{Rastinejad:25}, triangles from \citetalias{Li:25}, and circles from \citetalias{Gokul:25}). Pink, yellow, and cyan circles correspond to XRF-SN~2006aj, GRB-SN~2023pel, and SN~2020bvc, respectively. The black solid line represents our preferred model for SN~2025kg, which includes CSM interaction, a magnetar and some Ni. Black dashed and dash-dotted lines represent the Ni model and the magnetar model (see Sec. \ref{sec:results} for details). Pink and cyan lines correspond to models of SN~2006aj and SN~2020bvc \citep[and this work]{Roman:2023}, respectively. Error bars have been included when possible. Inset: A model with a different CSM distribution (dotted line) hints at a closer match at early-time velocities.}
    \label{fig:comparison}
\end{figure}

\section{Models and observations}\label{datos}

The bolometric LCs of SNe are very sensitive to their progenitor properties and power sources, while their early evolution can be strongly influenced by a nearby CSM. It is common to compare theoretical LC models with observations in order to derive the progenitor and explosion parameters. However, an important degeneracy arises when only photometric data are used. Including the photospheric velocity evolution, as inferred from some spectral lines, can help to break this degeneracy. Fe II velocities have been proposed as an effective tracer of photospheric velocity \citep{Dessart:2005}. Here, we used Fe II velocity measures by \citetalias{Gokul:25} to compare with our models.

Various authors have estimated the pseudo- or bolometric luminosity of SN 2025kg, and their results are overall in remarkable agreement (see black symbols in Fig. \ref{fig:comparison}). \citetalias{Gokul:25} produced a bolometric LC by applying the color-calibration method provided by \cite{Lyman:2014}; their data covered the early cooling phase and the main peak from 1 to 35 days. \citetalias{Li:25} computed a pseudo-bolometric LC through direct flux integration over rest-frame wavelengths of 3000-9000$\,\AA$. The data provided by \citetalias{Rastinejad:25} was calculated using grizJHK photometry and co-bands from ATLAS. Although the three LCs are similar around the main peak, significant discrepancies appear at early epochs \mbox{(t < 6 days)} during the first peak (cooling phase). It is possible that the non-inclusion of a UV correction accounts for these differences, even though such a correction carries high uncertainties. Accordingly, the earliest data—and parameters derived from them—should be treated with caution.

For modeling purposes, we adopt the bolometric LC of \citetalias{Gokul:25} and supplement it with the measurements of \citetalias{Li:25} to improve the temporal coverage. In addition, we include data from \citetalias{Rastinejad:25} as a comparison. Although the early data from the latter authors were not used in the modeling, the data point at $\sim$58 days was included, as it provides additional constraints on the synthesized nickel mass. The times refer to the detection of the XRF at UT 2025-01-08 12:30:28.34, which is taken as the explosion time throughout this paper.

We compare the observational data with the theoretical LC and photospheric velocity evolution calculated using a one-dimensional radiation hydrodynamic code \citep{Bersten:2011}. The code simulates the explosion by injecting some energy manually (E) near the core of the progenitor star (M$_{\rm cut}$), which we assume will collapse. This energy is responsible for the formation of a powerful shock wave that propagates inside the star, transforming thermal and kinetic energy into radiative energy. The code has a crude treatment of radiation transfer, assuming the diffusion approximation for optical photons and gray transfer for the gamma rays produced by the radioactive decay of $^{56}$Ni. Any nickel distribution is allowed in our code, and the gamma-ray deposition through the SN ejecta is calculated assuming a constant value for the gamma opacities of $\kappa_{\gamma}= 0.03 \ \rm{cm}^2\,\rm{g}^{-1}$ \citep{Sutherland:1984}. However, a detailed treatment is applied for the hydrodynamical variables, including relativistic effects that become important for fast-moving material. The code has been fully described in \cite{Bersten:2011} and has been used to model several SNe of different types, from H-rich to H-free objects \citep{Taddia:2018,Martinez:2022,Bersten:2024}, and from normal to more extreme SNe \citep{Bersten:2016, Gutierrez:2021,Orellana:2022}. Different energy sources can be included in the code in addition to the explosion energy and $^{56}$Ni decay, such as a magnetar source \citep{Orellana:2018} and/or the presence of some CSM \citep{Englert:2020,Ertini:2025}. 

As an initial input for our simulation, a pre-SN model representing the state of the star prior to the explosion is required. Hydrostatic structures, calculated with stellar evolution codes, are typically employed as pre-SN conditions. However, the outermost regions of these structures are often modified by hand when the effect of a CSM is incorporated, since no evolutionary models consistently reproduce the CSM conditions required to match the observations. 
In this work, we use the same grid of stellar models recently employed in \citet[hereafter RA25]{Roman:2025}. These models consist of H-free\footnote{He-free progenitors would be more appropriate for SNe Ic, but no self-consistent pre-SN models with complete He removal are currently available.} structures of different masses, specifically stars with main-sequence mass (M$_{\rm{ZAMS}}$) of 13, 15, 18, 20, and 25 M${_\odot}$ which correspond to pre-SN mass of 3.3, 4, 5, 6, and 8 M$_\odot$. All these models have a compact structure at the explosion time with R $\lesssim$ 5 $R_\odot$ before modifying its structure, considering the inclusion of CSM. 

We visually compared models and observations using luminosity and line-velocity measurements. For practical purposes, we first focus on reproducing the main peak, which is primarily powered by either $^{56}$Ni or a magnetar, and subsequently on the early-time emission. Note that this paper only considers the CSM interaction as a possible explanation for the first peak; however, different scenarios have been explored in previous works (see e.g., \citetalias{Gokul:25,Eyles-Ferris:25}).

\section{Results}\label{sec:results}

\begin{table*}[h]
    \centering
    \caption{Parameters derived through our hydrodynamic modeling for SN~2025kg, SN~2006aj, SN~2020bvc and SN~2023pel.}
    \resizebox{0.9\textwidth}{!}{
    \begin{tabular}{lccccccccccr}
    \hline
       SN & Model & E & M$_{\rm ZAMS}$ & M$_{\rm Ni}$ & M$_{\rm ej}$ & P & B & R$_{\rm CSM}$ & M$_{\rm CSM}$ & Reference   \\
        &  & [foe] & [M$_{\odot}$] & [M$_{\odot}$] & [M$_{\odot}$] & [ms] & [10$^{14}$ G] & [R$_{\odot}]$ & [M$_{\odot}$] & \\
        \hline
        2025kg & Ni & 2.2 & 13 & 0.85 & 1.9 & --- & --- & --- & --- & This work \\
         & Mag+CSM & 2.4 & 18 & 0.2 & 3.4 & 2.9 & 28 & 500 & 0.27 & This work \\
         \hline
        & & 2.91$^{+1.36}_{-0.86}$ & --- & 0.57$^{+0.6}_{-0.3}$ & 1.66$^{+0.79}_{-0.49}$ & --- & --- & $\sim 575$ & $ 0.07^{+0.06}_{-0.04}$ & \citetalias{Gokul:25} \\
        2025kg & Other & $\sim$14 & --- & --- & $2.42^{+0.67}_{-0.7}$ & 14.46$^{\pm 0.12}$ & 2.56$^{\pm 0.06}$ & --- & --- & \citetalias{Li:25} \\
        & Authors & --- & 15--30 & 0.2--0.6 & 0.8$^{+0.5}_{-0.2}$ & --- & --- & --- & --- & \citetalias{Rastinejad:25}\\
        & & $\lesssim$1 & --- & --- & --- & --- & --- & $\sim\,3000$ & 0.2--0.9 & \citetalias{Eyles-Ferris:25} \\
       \hline
       \hline
        2006aj & Ni+CSM & 3 & 13 & 0.28 & 1.1 & --- & --- & 200 & 0.024 & This work\\
        2020bvc & Ni+CSM & 7.5 & 15 & 0.23 & 2.6 & --- & --- & 100 & 0.04 & This work \\
        2023pel & Magnetar & 2.3 & 18 & 0.24 & 3.4 & 3.2 & 28 & --- & --- & \citetalias{Roman:2025} \\
        \hline
    \end{tabular}
    }
    \tablefoot{The parameters derived for SN~2025kg by other authors, using semi-analytical models, are also presented.}
    \label{tab:all_SNe_parameters}
\end{table*}

Figure~\ref{fig:comparison} compares SN~2025kg with other energetic SNe. SN~2006aj associated with an XRF \citep{Simon:2010,Pian:2006}, SN~2020bvc, a SN Ic-BL with some evidence of an offset jet \citep{Ho:2020}, and SN~2023pel, the most recent GRB-SN (\cite{Gokul:2024,Hussenot-Desenonges:2024}). From the figure, we highlight some aspects: (1) there is a clear diversity in luminosity within the XRF-SN objects, (2) early emission prior to the main peak is present in all the comparison events. Interestingly, this early emission appears to be more frequent in SNe associated with high-energy radiation than in normal SN Ic, and it may provide some insights into their progenitor origin. (3) Despite the large range of radiative output (luminosity), the kinetic energy (velocities) seems to exhibit a more homogeneous behavior, and (4) SN~2025kg is remarkably similar to SN~2023pel in terms of both luminosity and the Fe II velocities. 

Recently, \citetalias{Roman:2025} presented a detailed hydrodynamic model of SN~2023pel. In that work, we noted that SN~2023pel was brighter than most of GRB-SNe and exhibited relatively low expansion velocities. A magnetar central engine was proposed to account for these intrinsic properties. Given the similarity between both events, a similar explanation is expected to apply to SN~2025kg. 

Figure~\ref{fig:comparison} presents some of our preferred models for SN~2025kg; a model powered only by $^{56}$Ni (the Ni model) and a magnetar plus some $^{56}$Ni (the magnetar model, for simplicity). About $\sim$ 0.2 M$_\odot$ of nickel was also included in the latter to improve the agreement with the observations at later times (t $\sim 58$ days), and because some $^{56}$Ni is naturally expected to be synthesized during the explosion.  
Finally, our selected model consists of a magnetar plus a CSM (Mag+CSM). The parameters of each of these models are listed in Table \ref{tab:all_SNe_parameters}.

All the models presented in Fig.~\ref{fig:comparison} reproduce the data of SN~2025kg reasonably well around the main peak. Note that models including a magnetar contribution require a more massive progenitor (5 M$_\odot$; see Tab. \ref{tab:all_SNe_parameters}) than the Ni model. The same was previously noted in our analysis of SN~2023pel. This is because the magnetar model supplies additional energy, which results in a narrower LC. To counterbalance this effect and reproduce the observed LC width, a more massive progenitor is needed to achieve the necessary broadening.
The Fe II velocities are also well reproduced, except for the first data point, which none of our models could reproduce. It is possible that the measurement of this line velocity at early epochs is subject to large uncertainties.\footnote{Alternatively,  a different CSM distribution than a steady wind could help to improve the agreement (see the inset in Fig.~\ref{fig:comparison}).}. Here, we present only our favorite models, although several alternatives were previously explored to select an acceptable solution in each scenario. These calculations involved different progenitor masses and variations in the free model parameters (E, M$_{\rm{Ni}}$, and P and B). However, given the similarity between SN~2025kg and SN~2023pel, the exploration was guided by our previous results.

As mentioned before, both the Ni- and magnetar models provide a good representation of SN~2025kg observations. However, the Ni model requires a large amount of nickel mass (M$_{\rm{Ni}}=$ 0.85 M$_\odot$), especially considering the low ejecta mass of this model (M$_{\rm{ej}}=$ 1.9 M$_\odot$). This high nickel mass is required to account for the high luminosity observed in this object. This was precisely the main argument used by \citetalias{Roman:2025} to favor an additional energy source in the case of SN~2023pel, and it was also invoked by \cite{Bersten:2016} for SN~2011kl. In a similar way, we believe that this reasoning applies to SN~2025kg; therefore, we favor the magnetar scenario and adopt it to model the early-time emission.

In this work, we only explore the possibility that CSM interaction powers the early emission. To include the effect of the CSM, we attached some material in the outermost layer of the pre-SN density profile, assuming a stationary wind law ($\rho \propto r^{-2}$). After exploring various configurations, primarily adjusting the CSM extension and mass, we identified a model that accurately reflects the data. This model is presented in Fig.~\ref{fig:comparison} and has a CSM extension and mass of 500 R$_\odot$ and 0.27 M$_\odot$, respectively. A wind velocity of 115 $\rm{km}\, s^{-1}$ was assumed in our calculations. Although CSM models reproduce the initial LC well, they yield a poorer match to the first velocity data. In fact, our CSM models produce even lower velocities at those times. However, adopting a different CSM distribution--rather than the steady wind used here--could improve the velocity match without significantly affecting the initial LC behavior (see the inset in Fig.~\ref{fig:comparison}).

In Fig.~\ref{fig:comparison} we also present models for the SN~2006aj and SN~2020bvc calculated using the same code and progenitor grid. The parameters of the models are given in Tab.~\ref{tab:all_SNe_parameters}. In these cases, only a nickel power source was explored, due to the relatively normal luminosities. For modeling the early phase, we attached the CSM on top of the external density profile, as explained above for SN~2025kg. The values found for M$_{\rm Ni}$ are within the expected range for H-free SNe and are considerably lower than those found for SN~2025kg (Ni model), in agreement with their maximum luminosities. On the other hand, SN~2020bvc has the highest E of the sample, consistent with the behavior of its velocities, whose values remain higher throughout their evolution. SN~2006aj shows the lower M$_{\rm ZAMS}$ and M$_{\rm ej}$, in agreement with showing the narrowest LC. The properties of the CSM are also in concordance with the behavior shown in their LC, as a slower decay in the early phase is associated with higher values of M$_{\rm CSM}$ and R$_{\rm CSM}$\footnote{Although some degeneration between M$_{\rm CSM}$ and R$_{\rm CSM}$ exists.}, which in turn produces higher values of the LC minimum. Finally, we note that the model parameters of SN~2023pel are in very good agreement with the value obtained for our preferred model for SN~2025kg (Mag+CSM) which is expected given the similarities between both SNe.

\section{Comparison with previous works}\label{sec:disc}

Table~\ref{tab:all_SNe_parameters} shows the parameters derived by other authors for the SN~2025kg. As mentioned previously, analytical models were used in prior studies, relying only on photometric data; therefore, differences are to be expected. Despite this, we found good agreement between our Ni model and that of \citetalias{Gokul:25}, when the reported uncertainties are considered. All physical properties, and the parameters derived for the CSM, are generally consistent with our results. The agreement is less satisfactory when comparing our Ni model parameters with those derived by \citetalias{Rastinejad:25}. However, the authors do not provide an estimate for the explosion energy, which could have a non-negligible impact on the derived mass values. Regarding the CSM properties, the extent and mass values reported by \citetalias{Eyles-Ferris:25} are considerably larger than ours, particularly the value of R$_{\rm CSM}$. This is noteworthy because the data used in their analysis are systematically less luminous than those considered in this work (see \citetalias{Rastinejad:25} data in Fig. \ref{fig:comparison}). Therefore, one would have expected some differences in the opposite direction, with lower CSM extension and mass, as found for SN~2006aj and SN~2020bvc (see Tab~\ref{tab:all_SNe_parameters}). Although the low value of explosion energy (E $\lesssim$ 1 foe) assumed in \citetalias{Eyles-Ferris:25} could perhaps explain some of the differences, it is unlikely to be the only reason. 

\cite{Li:25} pointed out that SN~2025kg is a highly luminous event, and were the first to propose a magnetar as its potential energy source, as we suggest in this work. However, the inferred parameters differ significantly from those found here, with only the M$_{\rm ej}$ value being broadly consistent with our results. The E value is significantly higher and it was unclear for us which spectroscopic data were used to infer it. Furthermore, with their adopted values of M$_{\rm ej}$ and E, it appears challenging to explain the evolution of the expansion velocities of SN~2025kg, which are not particularly large. In addition, the values of P and B differ significantly from those used in our modeling. We have tested models using the parameter values presented by \citetalias{Li:25}, but we were unable to reproduce the observations of this SN. The resulting model shows noticeable deviations from the data. However, we were unable to identify the reason for these discrepancies.

\section{Conclusions}\label{sec:conclus}

The luminous SN~2025kg is another example from the small group of SNe accompanied by an XRF. Its LC exhibits two components: an early cooling emission and a main peak. The main peak and the expansion velocities are remarkably similar to those observed in SN~2023pel, which was associated with a GRB.

Our numerical models indicate that a large amount of nickel (M$_{\rm{Ni}}\sim$ 0.85 M$_\odot$) and a low ejecta mass (M$_{\rm{ej}}\sim$ 1.9 M$_\odot$) are required to explain the observations when considering a model powered only by $^{56}$Ni. Alternatively, a model including an additional energy source provided by a magnetar—with \mbox{P = 2.9 ms} and \mbox{B = $28 \times 10^{14}$ G}—and a typical nickel mass ($\sim$0.2 M$_\odot$), can also reproduce the observations. As in the case of SN~2023pel, we also favor the magnetar scenario for SN~2025kg, as we think this model is physically more plausible (see \citetalias{Roman:2025}). On the other hand, the early LC component was modeled by assuming the presence of some CSM located near the progenitor star before the explosion. By adopting an extension of 500 R$_\odot$ and a CSM mass of 0.27 M$_\odot$, we were able to reproduce the early-time emission under the assumption of a steady wind profile. These parameters imply an extreme mass-loss episode $\sim$ 0.1 yr before the explosion, which may be difficult to justify physically. An alternative is that the progenitor had an extended envelope, as that needed to reproduce early data in SN IIb \citep{Bersten:2012}. Such an envelope could naturally arise from binary mass transfer, which could also favor the magnetar formation scenario. 

When comparing the properties of SN 2025kg with those of other events associated with high-energy radiation, we find that their luminosities are highly diverse, while their expansion velocities exhibit much smaller variations. Conversely, an early emission component appears to be relatively common among these objects, unlike in most normal He-deficient SNe. Some of the observed characteristics seem easier to explain within a binary evolution scenario. We therefore speculate that binary progenitors are likely responsible for this class of high-energy transients.

\begin{acknowledgements}
M.B. acknowledges support from PIP 112-202001-10034 and PICT 2020-1141 grants and to CSIC-iCOOP exchange program. 
\end{acknowledgements}

%
   \bibliographystyle{aa} 
   \bibliography{biblio} 
%





\end{document}